\documentclass[5p]{elsarticle} 
\usepackage{graphicx}
\usepackage{amssymb,amsmath,amsfonts,latexsym}
\usepackage{booktabs}
\usepackage{multirow}
\usepackage{rotating}
\usepackage{float}
\usepackage{fancyhdr}
\def\gs{\Upsilon_\star}

\begin{document}

\title{Testing Grumiller's modified gravity at galactic scales}
\author[rvt]{Jorge L. Cervantes-Cota\corref{cor1}} 
\ead{jorge.cervantes@inin.gob.mx} 
\cortext[cor1]{Corresponding author} 
\author[rvt]{Jes\'us A. G\'omez-L\'opez}

\address[rvt]{Depto. de F\'{\i}sica, Instituto Nacional de
Investigaciones Nucleares, Apartado Postal 18-1027 Col. Escand\'on, 11801 DF, M\'{e}xico.} 

\begin{abstract}
Using galactic rotation curves, we test a -quantum motivated- gravity model that at large distances modifies the Newtonian potential when spherical 
symmetry is considered. In this model one adds a Rindler acceleration term to the 
rotation curves of disk galaxies. Here we consider a standard and a power-law generalization of the Rindler modified Newtonian potential that are 
hypothesized to play the role of dark matter in galaxies. The new, universal  acceleration has to be -phenomenologically- determined.   Our 
galactic model includes the mass of the integrated gas and stars for which we consider a  free mass model.  We test the model by fitting rotation 
curves  of thirty galaxies that has been employed to test other alternative gravity models.    We find that the Rindler parameters do not perform 
a suitable fit  to the rotation curves in comparison to the Burkert dark matter profile, but the models achieve a similar fit as the NFW's profile does. However, the computed 
parameters of the Rindler gravity show some spread, posing the model 
to be unable to consistently explain the observed rotation curves. 
\end{abstract}

\maketitle
\section{Introduction}
It is well known that General Relativity  is a theory well tested within the solar system and scales below, inasmuch as no deviations 
to it have been found since many years \cite{Fischbach:1999bc,Will:2005va}. However, new theories/models of 
gravitation have been recently  proposed motivated by different theoretical and observational reasons, see Ref.  \cite{Clifton:2011jh} for a  review. 
One of the motivations is to test gravity theories beyond the solar system, and to understand what constraints could be drawn at 
different length scales.  On the one hand, at cosmological scales different corrections apply to the standard theory of large scale 
structure alone from General Relativity \cite{Yoo:2009au,Yoo:2010ni,Jeong:2011as}  and, in addition, new approaches have been put forward 
to understand  the possible deviations of data to the theory \cite{Hu:2007pj,Pogosian:2010tj,Daniel:2010ky,Lombriser:2013aj}.   On the other hand, at galactic scales 
rotation curves provide a unique laboratory to test kinematical deviations from 
theoretical expectations  
and in fact rotation curves are one of the reasons why dark matter has been hypothesized. 
Although cold dark matter is the most popular candidate, there are other possibilities, e.g. bound dark matter  \cite{delaMacorra:2009yb,delaMacorra:2011df,Mastache:2011cn}, or other theoretical approaches that modify gravity or kinematical laws such as MOND  \cite{MOND1,MOND2}  (see however \cite{Cardone:2012qy}) or  f(R)-gravity 
that apart from playing the role of dark energy also intends to replace dark matter \cite{Capozziello:2004us,Martins:2007uf} .   

Recently, a model of gravity has been put forward that stems from quantum gravity corrections to General Relativity and when one applies it to spherical symmetry 
and local (galactic) scales an extra Rindler acceleration appears in addition to the standard Newtonian formula for 
rotation curves \cite{Grumiller:2010bz,Grumiller:2011gg}. The 
new Rindler term is hypothesized to play the role of dark matter in galaxies.  This idea has been tested already in a very recent work 
\cite{Lin:2012zh}, where a fit is made to eight galaxies of the The HI Nearby Galaxy Survey (THINGS) \cite{Walter:2008wy}. They found that 
six of the galaxies tend to fit well to the data and that there is a preferred Rindler acceleration parameter of around 
$a \approx 3.0 \times 10^{-9} \, {\rm cm}/{\rm s}^2$ ($=926 \, {\rm km}^{2}/{\rm s}^2 {\rm kpc}$); they later fixed this acceleration parameter and found acceptable fits for five galaxies, and 
furthermore, an additional free parameter let them to fit two more galaxies.  
We have revised this idea using a greater sample (seventeen) of THINGS galaxies, and for the eight original galaxies we find similar 
conclusions on the fits and to a convergence to a similar Rindler acceleration within $1 \sigma$ confidence level.   But  when 
one adds more  galaxies to the analysis the spread in the acceleration blows up, and therefore we concluded that the model is not tenable \cite{Mastache:2012ep}.   
However,  THINGS rotation curves are based on gas kinematics, whereas there are claims pointing out that complex gas dynamics could not be 
a good tracer of gravity in spirals, and  gas and stellar motion do not exactly coincide in all the cases \cite{Gomez:2013hz}. Given this,  
in the present work we test again the Rindler acceleration hypothesis (and a generalized version of it) but with a different sample of galaxies that is larger (thirty galaxies) than the previous 
sample and has very different systematics.  This set of galaxies has been used to test other gravity models in the past \cite{Martins:2007uf} since 
most of them  have desired properties such as smoothness in the data, symmetry, and they are extended to large radii.
   
This work is organized as follows: In Section \ref{grumiller:eqs} we briefly review Grumiller's model of gravity at large distances,  in Section \ref{grumiller:rc:model} we explain the 
rotation curve models, in Section  \ref{grumiller:results} we present our results and compare the fits to results from standard dark matter profiles such as Navarro-Frenk-White (NFW) \cite{Navarro:1995iw,Navarro:1996gj} and Burkert  \cite{Burkert:1995yz}. Finally, Section \ref{grumiller:conclu} is devoted to conclusions. Supplementary material is included to 
support our conclusions.   

\section{Grumiller's gravity model at large distances} \label{grumiller:eqs}  
In order to have a self-consistent description of this work we briefly review the main ideas behind 
Grumiller's model, for details see  \cite{Grumiller:2010bz}.  The model starts 
with spherical symmetry in four dimensions split in the following way:    
 \begin{equation}\label{curvatura}
ds^2=g_{\alpha\beta} dx^{\alpha} dx^{\beta} + \Phi^2 (d\theta^2+ {\rm sen}^2 \theta d \phi^2)      \, , 
\end{equation}
where $g_{\alpha\beta}(x^{\gamma})$  is a 2-dimensional metric and the surface radius $\Phi(x^{\gamma})$ 
depend upon $x^{\gamma}=\{t,r\}$.  The idea is to describe  
these fields in two dimensions since the gravitational potentials $g_{\alpha\beta}$ and $ \Phi$ that are 
intrinsically two-dimensional, and their solutions can be mapped 
into the 4-dimensional world through Eq. (\ref{curvatura}). 

The most general 2-dimensional gravitational theory that is renormalizable, that yields a standard Newtonian potential, and 
that avoids curvature singularities at large $\Phi$ is:  

\begin{equation}\label{accio} 
S=-\int\sqrt{-g}[\Phi^{2} R+2{\partial\Phi}^2-6\Lambda\Phi^2+8a\Phi+2]d{^2x}  \, , 
\end{equation}
that depends on two fundamental constants, $\Lambda$ and  $a$, the cosmological constant and a Rindler acceleration, respectively. The solutions to this action 
will describe the original line element, Eq. (\ref{curvatura}), that will model gravity in the infrared.   The solutions are:  
\begin{equation}
\label{result1} 
g_{\alpha\beta}dx^{\alpha}dx^{\beta} = - K^{2} dt^{2} +  \frac{dr^2}{K^2} ,
\end{equation}
\begin{equation}
\label{result2} 
K^{2} =1- \frac{2M}{r}- \Lambda r^{2} + 2 a r, 
\end{equation}
with $K$ being the norm the Killing vector $\partial_t$ and $M$ a constant of motion.  Of course, if $\Lambda= a = 0$, one recovers the Schwarzschild solution.  If 
$M= \Lambda = 0$, it yields the 2-dimensional Rindler metric. Therefore, the resulting gravity theory differs from General Relativity only by the addition of a Rindler 
acceleration, see also \cite{Grumiller:2002nm}.     

A geodesics study of time-like  test particles moving in a 4-dimensional spherical, symmetric background, according to Eqs. (\ref{curvatura}), 
(\ref{result1}), and (\ref{result2}),  results in the following eqs:
\begin{eqnarray} \label{eqs_motion}
E &=&  \frac{\dot r^2}{2}+ V_{\rm eff}  \\
V_{\rm eff} &=& -\frac{M}{r} + \frac{\ell^2}{2r^2}  + ar\,\big(1+\frac{\ell^2}{r^2}\big) , 
\end{eqnarray}
where $E=$ const. and $V^{\rm eff} $ is the effective potential.

When we apply the previous solution to a galactic are\-na, we set the cosmological constant equal to zero ($\Lambda =0$) since the mean energy density of a galaxy 
is much larger than the cosmic inferred $\Lambda$.   We set also $l=0$, to avoid an additional angular momentum to the system that in fact shall not account for the 
kinematical deficit of rotational curves.  

Considering now the effects on rotation curves, the Rindler acceleration yields an additional term in the rotation's speed ($v_T$):      
\begin{equation}
\label{velgru1} 
v_{T}(r)=\sqrt{r\left| \frac{d\phi_{T}}{dr} \right|+ar} ,
\end{equation}
where $\phi_{T}$ is the total gravitational potential that  test particles (stars and gas) feel. This is the original Grumiller's model of gravity at large 
distances \cite{Grumiller:2010bz}. The new Rinder acceleration term should account for the kinematical difference of the observed and 
predicted rotation curves.  Notice that Eq. (\ref{velgru1}) 
diverges asymptotically, at large radius. This is not an observed behaviour in typical rotation curves, but on the contrary they tend to slowly decrease after a 
few optical radii  \cite{Salucci:2007tm}.  Therefore, as a generalization of the previous model one may intend to determine a power-law dependence in the Rindler term, 
as suggested in Ref. \cite{Grumiller:2010bz}.  The new term should not diverge at large distances.   Accordingly, we will consider the following  
{\it generalized} Grumiller model: 

\begin{equation}
\label{velgru2} 
v_{c_{T}}(r)=\sqrt{ r \left| \frac{d\phi_{T}}{dr}  \right|+a \, r^n} ,
\end{equation}
where there are two undetermined Rindler parameters $( a, n )$. The case $n=1$ yields acceleration units to $a$, but a different $n$ implies 
$\frac{{\rm length}^{2-n}}{{\rm time}^2}$ units; one could extract an acceleration parameter here if one defines 
$a \, r^{n} \equiv a_{\rm new} r \,  (r/r_{\rm new})^{n-1}$, but we would only add an extra parameter ($r_{\rm new}$) that  is completely 
degenerated with $a_{\rm new}$.   This could be done a posteriori, if needed.  


\section{Rotation curve model}  \label{grumiller:rc:model}
In this section we closely follow the model presented in Ref. \cite{Mastache:2012ep}, but for the sake of completeness we present it here again.    
The galaxy model consists of gas and stars orbiting on a disk plane, and instead of dark matter we include the Rindler acceleration, explained 
in the previous section.  The contribution of gas is computed by integrating the surface brightness as in the standard Newtonian case by  
assuming an infinitely thin disk. One directly integrates its contribution to the rotation curve ($v_G$).  

For stars we take a standard Freeman disk   
\cite{Fr70,BaSo80}: 
\begin{equation} \label{rho-stars}
  \rho_{\star} (r) = \frac{M_d}{2 \pi r_{d}^{2}}  \, e^{-r/r_{d}} ,
\end{equation}
where $M_d$ is the mass of the 
disk and  $r_d$ its radius.   The rotation curve contribution from stars within standard Newtonian dynamics, yields \cite{BiTr08}:
\begin{eqnarray}\label{v2_stars}
 \hspace{-0.5cm}  v_{\star}^{2}(r)  = \frac{G M_d }{2 r_{d}} \left(\frac{r}{ r_{d}}\right)^{2}  && \left[ 
   {\rm I_{0}}\left(\frac{r}{2 r_{d}}\right) {\rm K_{0}} \left(\frac{r}{2 r_{d}}\right)   \right. \nonumber \\ 
  &&  \left. -{\rm I_{1}}\left(\frac{r}{2 r_{d}}\right) {\rm K_{1}} \left(\frac{r}{2 r_{d}}\right) \right] ,
\end{eqnarray}
where I and K are the modified Bessel functions. 

The stars' contribution to the rotation curves is normally multiplied by the mass-to-light ratio ($\gs$), that is an additional free parameter
in the mass model, introduced because we generally can only measure the distribution of the light instead of the mass.
When we estimate the Rindler parameters $(a,n)$, $\gs$ is an important source of uncertainty, because these parameters are degenerate 
through Eq. (\ref{vel2}), see below. However, 
since stars have a major contribution near the center of the galaxy and the Rindler acceleration contribute most at large distance, $\gs$ does not 
significantly affect  the uncertainties of the Rindler parameters, as we have shown in Ref. \cite{Mastache:2012ep}. 

The $\gs$ has been modeled, e.g. in Salpeter \cite{Salpeter:1955it}, Kroupa \cite{Kroupa:2000iv}, and Bottema \cite{Bottema:1997qe}, but the 
precise value for an individual galaxy is not well known and depends on extinction, star formation history, initial mass function, among
others. Some assumptions have to be made respect to $\gs$ in order to reduce the number of free parameters in the model.  In a 
previous work \cite{Mastache:2012ep}, one of us (JLCC) has studied the Kroupa, diet-Salpeter, and $\gs$ as a model-independent free parameter. It was shown that 
the different stellar mass models do not significantly change the determined value of the Rindler parameters for most of the 
galaxies, and  from the  three models, the free $\gs$ yields best fits to rotation curves.  Thus, for the purpose of this letter, we adopt the $\gs$ free mass model.

Gathering all contributions to the total ($T$) rotation curve
and including a generalized Rindler (GR) term\footnote{Notice that the squared sum in Eq. (\ref{vel2}) is meant to add the different 
gravitational contributions of gas, stars, and Rinder, but it is not meant to represent a vectorial's squared sum since the different velocities 
are not orthogonal contributions to the total. We thank Marcelo Salgado for pointing it out  about this commonly used notation.},
\begin{equation}\label{vel2}
    v_{T}^2 (r) = \gs v^2_{\star} + v^2_{G} + v_{GR}^{2}(r) \, ,
\end{equation}
where we explicitly use $\gs$ and therefore assume $M_d$ with a solar mass-to-light in Eqs. (\ref{rho-stars}, \ref{v2_stars}); the power-law generalized Rindler term is 
\begin{equation}\label{aRind.gen}
 v_{GR}^{2}(r) \equiv a |\vec{r}|^n .
\end{equation}
The case $n=1$ is the original model of modified gravity at large distances \cite{Grumiller:2010bz,Grumiller:2011gg}, as the Rindler contribution in 
Eq. (\ref{velgru1}).  The new free parameters of the model of galactic rotation curves are $a$ and $n$, and they have to be determined by observations. 
In standard dark matter profiles such as NFW \cite{Navarro:1995iw,Navarro:1996gj}, 
Burkert  \cite{Burkert:1995yz}, pseudo-isothermal, or alternative Bound Dark Matter \cite{delaMacorra:2009yb} one also uses two free parameters, 
and therefore the number of degrees of freedom to fit is same as in the generalized Rindler model; 
for a comparison of these profiles
see  Refs. \cite{delaMacorra:2011df,Mastache:2011cn}.   To extract information for the Rindler parameters, as an 
input we will need the observational rotation curve and the computed gas contribution. 


\section{Rotation curve fits and results} \label{grumiller:results}

To perform the fits we employed the same method as in Ref. \cite{Mastache:2012ep}, but here it is applied to a set of 
thirty galaxies that has been used in the past to test alternative gravity models \cite{Martins:2007uf}.  Most of galaxies 
possess wanted properties such as smoothness in the data, symmetry, and they are extended to large radii.  We fit the observational 
velocity curve to the theoretical model (\ref{vel2}) using the $\chi^2$ goodness-of-fit test ($\chi^2$ test), that computes the parameters' best fits.
In general the $\chi^2$ test statistics are of the form:
\begin{equation} \label{chi2}
   \chi^2 = \sum_{i=1}^{m} \left(\frac{v_{{\rm obs}_i}-v_{{\rm model}_i}(r, a, n)}{\sigma_i}\right)^2,
\end{equation}
where $\sigma$ is the standard deviation, and $m$ is the number of observations.  One defines the 
reduced $\chi^{2}_{\rm red} \equiv \chi^2/(m-p-1)$, in which $m$ is the number of observations and $p$ is the number of fitted parameters.  
The total velocity (\ref{vel2}) defines our model - $v_{\rm model}$ in Eq. (\ref{chi2})- and depends on the three parameters: $\gs$ (or alternatively 
the mass of the disk $M_d$ that we actually fit), and the two Rindler parameters $(a,n)$.  

We firstly analyze the original Rindler model ($n=1$) and proceed to fit the parameters $a$ and $M_d$. Their physical units are  
${{\rm km}^2/{\rm s}^2 {\rm kpc}}$ and $M_{\odot}$, respectively.  
We assume a flat prior for these 
parameters in the following intervals:   $10^7<M_d<10^{12}$ and $0<a<10000$.  Given the large space that would require 
to show all rotation curve fits and parameters' contour plots, we include this information as supplementary material.  We gather our results in Table \ref{grumiller1}. 
The uncertainties in the rotation velocity are reflected in the uncertainties in the model parameters.   
One observes some spread in the values for $a$, ranging from   ${341.26}_{-64.46}^{+64.84}$ for M31	
to $2891.25_{-287.25}^{+293.75}$ for NGC7339 to account for a difference of an order of magnitude, but the uncertainties are small to account  
for such a difference.  In addition to this discrepancy,  the fits to some of the galaxies present very high  $\chi^2_{\rm red}$ values that result in poor fittings.  
Only thirteen (of thirty) galaxies had $\chi^2_{\rm red} \le 1$, and for these later galaxies we have included a distribution plot (as supplementary material) 
that shows a big spread in the values of the acceleration parameter.  We also plot the B-band mass-to-light ratios that are similar to others reported in the literature 
\cite{deBlok:1996gs,Martins:2007uf}.

For the generalized model ($n\neq 1$) we consider a flat prior in the interval $0 < n < 10$, and the same other conditions as previous model.  We 
determine now the two Rindler parameters $(a, n)$ and the stellar disk's mass. The results are shown in Table  \ref{grumiller2}.   Again a spread is observed 
in the acceleration parameter, ranging from  ${0.26}_{-0.21}^{+1.09}$ for M31 to $11605.40_{-555.00}^{+564.60}$ for UGC 10981 resulting in a  
difference of four orders of magnitude.  Since the $a$ value for M31 is very small and is the only one with a value much  less than one hundred, and 
since we did not include gas data to the analysis, we may exclude it for this analysis. The second lowest value of $a$ is  $113.39 _{-85.89}^{+195.61}$ for UGC 6917. Still 
there is a difference of two orders of magnitude between the smallest and  biggest values for the Rindler acceleration.   On the other hand, the 
power-law exponent ranges from $0.16 \pm 0.02$ for NGC 6503 to  $3.31 \pm 0.54$ for M31 or, again excluding M31, $1.60\pm 0.03$  for IC 2574, which 
is an order of magnitude difference.  For both parameters the uncertainties are too small to account for the encountered differences.  Similarly as the 
previous Rindler model, one only has thirteen (of thirty) galaxies with $\chi^2_{\rm red} \le 1$, and again, for these later galaxies we have included 
distribution plots (as supplementary material) that show a big spread in both parameters of the generalized Rindler model.

\begin{table*}[t]
\begin{center}
\begin{tabular}{  c  c  c | c  c c c  } \hline \hline 
\bf Galaxy & \bf Type &  $R_d$ (kpc)  &  $a$ (${{\rm km}^2/{\rm s}^2 {\rm kpc}}$) &  $M_d$ ($M_\odot$) & $\Gamma_{\odot}^{B}$ &$\chi^2_{\rm red}$ \\ \hline 
DDO 47	&IB	&0.50&$	820  _{-50}^{+70}	$&$1.0_{-0.1}^{+0.4} \times 10^7 $&0.1&4.9\\ \hline
ESO 116-G12	&SBcd	&1.70&$	1010 \pm 50$&$	3.0 \pm0.4 \times 10^9$	&0.6&3.8\\ \hline
ESO 287-G13	&SBc	&3.28&$ 920 \pm 30$&$3.9_{-0.1}^{+0.2} \times 10^{10}$&1.3&1.8\\ \hline
IC 2574	&SABm&1.78&$	400 _{-10}^{+20} $&$1.0_{-1.0}^{+0.2} \times 10^7	$&$<$ 0.1&38.9\\ \hline
M 31$^{*}$	&Sb	&4.50&$	340 \pm 60	$&$	1.68_{-0.01}^{+0.02} \times 10^{11}	$&8.4&1.6\\ \hline
M 33		&Sc	&1.42&$	900 \pm 30	$&$	5.0\pm 0.1  \times 10^9	$&0.9&4.3\\ \hline
NGC 55	&SBm	&1.60&$	870 \pm 50	$&$	5.9 \pm 2.3 \times 10^8$&0.2&	3.7\\ \hline
NGC 300	&Scd	&1.70&$ 790 _{-30}^{+50}$&$	2.7 \pm 0.3 \times 10^9	$&1.2&0.7\\ \hline
NGC 1090	&Sbc	&3.40&$	580 \pm 20$&$	5.0 \pm 0.2 \times 10^{10}	$&1.3&3.8\\ \hline
NGC 2403$^{*}$	&Sc	&2.08&$	900 \pm 20$&$	1.4_{-0.3}^{+0.4} \times 10^{10}	$&1.7&4.0\\ \hline
NGC 3877	&Sc	&2.80&$1500 \pm 400$&$2.7\pm 0.6 \times 10^{10}$&1.0&	0.9\\ \hline
NGC 3917	&Scd	&3.10&$	1000 \pm 100$&$1.3\pm0.2 \times 10^{10}$& 1.2&3.6\\ \hline
NGC 3949	&Sbc	&1.70&$	2300 \pm 1200$&$1.5 \pm 0.5 \times 10^{10}$&0.8&	0.4\\ \hline
NGC 3953	&SBbc	&3.80&$	1300 \pm 400$&$8.7 \pm  1.2  \times 10^{10}$&2.1&	0.4\\ \hline
NGC 3972	&Sbc	&2.00	&$1800 \pm 300	$&$3.8_{-2.0}^{+2.1} \times 10^9	$&0.6&0.5\\ \hline
NGC 4085	&Sc	&1.60&$2700 \pm 600$&$	3.4_{-2.7}^{+2.8} \times 10^9	$&0.5&2.0\\ \hline
NGC 4100	&Sbc	&3.37&$	500 \pm 100$&$	6.7 \pm  0.4 \times 10^{10}	$&2.7&1.0\\ \hline
NGC 4157	&Sb	&2.60&	$940 \pm 100	$&$5.2 \pm 0.5 \times 10^{10}	$&1.7&0.8\\ \hline
NGC 4183	&Scd	&3.20&	$370 \pm 70 $&$1.6\pm0.2 \times 10^{10}	$&1.7&0.1\\ \hline
NGC 4217	&Sb	&2.90&	$1100 \pm 200	$&$4.6_{-0.6}^{+0.5} \times 10^{10}	$&2.2&0.7\\ \hline
NGC 5585	&SABc	&1.26&$	870 \pm 30	$&$1.3 \pm 0.1 \times 10^9	$&0.9&7.3\\ \hline
NGC 6503$^{*}$	&Sc	&1.74&$	610 \pm 10$&$1.37\pm 0.03 \times 10^{10}	$&2.7&6.8\\ \hline
NGC 7339$^{*}$	&SABb	&1.50&$	2900 \pm 300	$&$1.2 \pm 0.1 \times 10^{10}	$&1.6&1.4\\ \hline
UGC 128	         &Sd	&6.40&	$350 \pm 70$&$	2.6_{-0.7}^{+0.8} \times 10^{10}	$&3.0&0.2\\ \hline
UGC 6399	&Sm	&2.40&	$660 \pm 260$&$3.6 \pm  2.2 \times 10^9	$&2.3&0.3\\ \hline
UGC 6917	&SBd	&2.90&	$590 \pm 170 $&$1.0\pm  0.2  \times 10^{10}	$&2.3&0.2\\ \hline
UGC 6983	&SBcd	&2.70&	$510 \pm 100 $&$	1.1\pm0.2 \times 10^{10}$&2.7&	0.7\\ \hline
UGC 8017$^{*}$	&Sab	&2.10&	$2000 \pm 150	$&$1.58 \pm 0.04 \times 10^{11}	$&4.0&6.3\\ \hline
UGC 10981$^{*}$	&Sbc	&5.40&	$430 \pm 40	$&$2.10\pm 0.02 \times 10^{11}	$&1.8&16.0\\ \hline
UGC 11455$^{*}$	&Sc	&5.30&	$2800 \pm 200	$&$1.2\pm0.1 \times 10^{11}	$&2.6&17.7\\ \hline \hline

\end{tabular}
\\
${}^{*}$ For these galaxies we have no gas data.   
\end{center}
\caption{Best fits for the standard Rindler model ($n=1$). It is shown in column (2) the galactic type and in (3) its disk radius, in (4) the 
acceleration parameter, in (5) the galactic disk mass, (6) the B-band mass-to-light ratio  in solar units, and in (7) the $\chi_{\rm red}^{2}$.}  \label{grumiller1}
\end{table*}

By comparing both fits ($n=1$ vs $n \neq 1$), the goodness of fits are better in the generalized model for sixteen galaxies, and for fourteen both models are 
equally well fitted. The Rindler acceleration varied 
for the standard Rindler model one order of magnitude and  for the generalized model two orders of magnitude. In our previous work, when one of us 
analyzed  the THINGS' galaxies \cite{Mastache:2012ep}, both the $\chi^{2}_{\rm red}$ and Rindler acceleration values changed more substantially: two and three 
orders of magnitude, respectively.   With respect to the power-law exponent of the generalized model, the present analysis results in one order of magnitude 
difference, whereas for the THINGS  galaxies resulted in two orders of magnitude.  The reason to have smaller differences in the computed parameters 
is that the present set of galaxies, while it is a larger collection, its uncertainties in data are bigger.  Although the present analysis soften 
the difference in the parameter computation, still such differences are not justified.  On the other hand, the mass-to-light ratios  found 
do not present systematic differences. 

From the set of galaxies considered in the present work and that in Ref. \cite{Mastache:2012ep} there is a 
common galaxy, NGC 2403.  The data considered are different and subject to different systematics. However, we 
may expect some similar parameter estimation.   The computed parameters in Ref. \cite{Mastache:2012ep} were, for the $n=1$ Rindler model,  
$a = 797.22^{+97.65}_{-0.32}$ and $M_{D} = 10^{10.2^{+9.8}_{-6.4}}$ $M_\odot$, whereas for the present computation table \ref{grumiller1} shows $a = 900 \pm 20$ 
and $M_{D} = 1.4^{+0.4}_{-0.3} \times 10^{10}$ $M_\odot$.  Clearly, both Rindler parameters and stellar disk masses are within 1 $\sigma$.  For the generalized model
the previous work gives $a = 3070 \pm 16$, $n=0.59 \pm 0.002$, and $M_{D} = 10^{9.9\pm 8}$ $M_\odot$, whereas in the present work we have $a = 3600 \pm 200$,   
$n=0.54 \pm 0.02$,  and $M_{D} = 8 \pm 0.4 \times 10^{9}$ $M_\odot$. In this case, the Rindler parameters are not quite different, but given the uncertainties, they are 
a few $\sigma$ away from each other;  stellar disk masses are within 1 $\sigma$.  
The discrepancy in Rindler parameters' uncertainties may be due to the fact that the THINGS sample include more data and are more precise. 
On the other hand, we do not expect a big influence of a small bulge, that was taken into account in  Ref. \cite{Mastache:2012ep},  
on the computed values of the Rindler parameters, since the main influence of the modified gravity is in the outer parts of the galaxies, where the bulge, or even 
the disk, counts less.    In our present work, we have not taken 
into account bulge contributions, since it is known the present set has negligible bulges \cite{Martins:2007uf}. 

\begin{table*}[t]
\begin{center}
\begin{tabular}{  c   | c  c  c  c  c} \hline \hline
\bf Galaxy &  $a$ (${{\rm km}^2/{\rm s}^2 {\rm kpc}^{n}}$) & $n$ &  $M_d$ ($M_\odot$) & $\Gamma_{\odot}^{B}$ &$\chi^2_{\rm red}$ \\ \hline 
DDO 47	&$	420 \pm 60	$&$1.5\pm 0.1	$&$3.3_{-2.1}^{+2.3} \times 10^{7} $ & 0.3&1.2 \\ \hline
ESO 116-G12	&$	1100 \pm 200$&$	1.0 \pm 0.1$ & $2.7 \pm 0.4 \times 10^9$&0.6&3.7\\ \hline
ESO 287-G13	&$	320_{-70}^{+90}$&$	1.3 \pm 0.1	$&$4.3 \pm 0.1 \times 10^{10}	$&1.5&1.6\\ \hline
IC 2574	&$ 180 \pm 8	$&$1.6\pm0.0$&$1.1 \pm 0.2 \times 10^8	$&0.1&2.4\\ \hline
M 31		&$	{0.3}_{-0.2}^{+1.1} $&$3.3 \pm 0.5$&$	1.8 \pm 0.0 \times 10^{11}	$&8.8&1.1\\ \hline
M 33		&$	1600\pm 100	$&$0.8 \pm 0.0$&$	3.8\pm 0.1 \times 10^9	$&0.7&3.3\\ \hline
NGC 55	&$	1000 \pm100	$&$0.9 \pm 0.1$&$	3.9_{-1.8}^{+1.5} \times 10^8$&0.1&	3.5\\ \hline
NGC 300	&$	630_{-120}^{+130}$&$	1.1 \pm 0.1 $&$	3.0 \pm 0.3 \times 10^9	$&1.3&0.7\\ \hline
NGC 1090	&$	1400 \pm 300$&$	0.8 \pm 0.1 $&$	4.5 \pm 0.2 \times 10^{10}	$&1.2&3.4\\ \hline
NGC 2403	&$	3600 \pm 200$&$	0.5\pm 0.0$&$	8.0 \pm 0.4 \times 10^9	$&1.0&2.3\\ \hline
NGC 3877	&$	550_{-340}^{+640}$&$	1.4 \pm 0.4 	$&$3.2 \pm 0.4 \times 10^{10}$&1.2&	0.9\\ \hline
NGC 3917	&$	430_{-140}^{+180}$&$	1.3_{-0.1}^{+0.2}	$&$1.7 \pm 0.2 \times 10^{10}$&1.5&	3.4\\ \hline
NGC 3949	&$	800_{-590}^{+1160}$&$	1.5_{-0.6}^{+0.8}	$&$1.8_{-0.2}^{+0.3} \times 10^{10}$&0.9&	0.4\\ \hline
NGC 3953	&$	7900_{-3100}^{+4500}$&$	0.5 \pm 0.2	$&$6.0_{-1.2}^{+1.3} \times 10^{10}$&1.5&	0.4\\ \hline
NGC 3972	&$    880_{-350}^{+460}	$&$1.3_{-0.2}^{+0.3}	$&$6.4_{-1.4}^{+1.5} \times 10^9	$&1.0&0.5\\ \hline
NGC 4085	&$	1400_{-600}^{+900}$&$	1.3_{-0.3}^{+0.4}$&$	5.6_{-1.7}^{+1.8} \times 10^9	$&0.8&2.0\\ \hline
NGC 4100	&$	770_{-440}^{+730}$&$	0.9_{-0.2}^{+0.3}$&$	6.6 \pm 0.4 \times 10^{10}	$&2.6&1.0\\ \hline
NGC 4157	&	$2600_{-1100}^{+1700}	$&$0.7 \pm 0.2	$&$4.5 \pm 0.5 \times 10^{10}	$&1.5&0.7\\ \hline
NGC 4183	&	$830 _{-390}^{+580}	$&$0.8 \pm 0.2	$&$1.4 \pm 0.2  \times 10^{10}	$&1.4&0.1\\ \hline
NGC 4217	&	$730_{-410}^{+690}	$&$1.2 \pm 0.3	$&$4.8 \pm 0.4 \times 10^{10}	$&2.3&0.7\\ \hline
NGC 5585	&$	1000 \pm 100	$&$0.9_{- 0.0}^{+ 0.1}	$&$1.1 \pm 0.1  \times 10^9	$&0.7&6.9\\ \hline
NGC 6503	&$	7900 \pm 400	$&$0.2\pm 0.0	$&$4.3\pm 0.3 \times 10^9	$&0.9&1.4\\ \hline
NGC 7339	&$	2000 \pm 300 $&$1.2 \pm 0.1	$&$1.3 \pm 0.1 \times 10^{10}	$&1.8&1.4\\ \hline
UGC 128	         & $260_{-180}^{+320} $&$	1.1 \pm 0.3 $&$	2.8 \pm 0.6  \times 10^{10}	$&3.2&0.2\\ \hline
UGC 6399	&	$240_{-170}^{+340}$&$	1.4_{-0.5}^{+0.6}	$&$5.1_{-1.2}^{+1.3} \times 10^9	$&3.2&0.3\\ \hline
UGC 6917	&	$110 _{-90}^{+200}$&$	1.6_{-0.5}^{+0.7}	$&$1.3 \pm 0.1 \times 10^{10}	$&2.9&0.2\\ \hline
UGC 6983        &	$340_{-170}^{+280}$&$	1.1 \pm 0.3 $&$	1.2 \pm 0.1  \times 10^{10}$&2.9	&0.7\\ \hline
UGC 8017	&	$8500 \pm 500 	$&$0.6 \pm 0.0	$&$1.1\pm 0.0 \times 10^{11}	$&2.9&4.8\\ \hline
UGC 10981${}^{*}$ 	&	$12000 \pm 600	$&$0.2 \pm 0.0	$&$1.5 \pm 0.0 \times 10^{11}	$&1.2&10.4\\ \hline
UGC 11455	&	$650_{-120}^{+140}	$&$1.5\pm 0.1	$&$1.5\pm 0.0 \times 10^{11}	$&3.3&16.9 \\ \hline \hline
\end{tabular}
\\
${}^{*}$ The upper limit for $a$ has been extended to find the $\chi^2$ minimum. 
\end{center}
\caption{Best fits for the power-law generalized Rindler model. It is shown in column (2) the acceleration parameter, in (3) the power-law exponent, in (4) the galactic disk mass, (5) the B-band mass-to-light ratio  in solar units, and in (6)  the $\chi_{\rm red}^{2}$.} \label{grumiller2} 
\end{table*}

We now compare the Rindler models with standard dark matter profiles, such as NFW \cite{Navarro:1995iw,Navarro:1996gj} and Burkert \cite{Burkert:1995yz}. The 
former is an example of a cuspy dark matter profile, whereas the later is shallow.  The explicit  computations are as those done in our previous works 
\cite{Mastache:2011cn,Mastache:2012ep} and are included as supplementary material.  To compare among the different models we constructed Table \ref{chi2_comp} with  
the $\chi^{2}_{ \rm red}$ values for  NFW, Burkert,  standard Rindler with $n=1$, and generalized Rindler ($n$-free), for the free stellar mass model.   The results 
are as follows:

\begin{itemize}

\item As already mentioned, the Rindler model with two free parameters ($a$, $n$) fits equally well or better than the model with a single parameter ($a$, $n=1$) for all galaxies. 

\item The standard Rindler model ($n=1$) fits worst than Burkert's profile, but similarly well as NFW. The standard Rindler model achieves an equally well or a 
better fit than both NFW and Burkert only for six galaxies (M 31, NGC 3949, NGC 3953, NGC 4183, NGC 4217, and UGC 6917) and, in addition, it fits better than 
Burkert for one galaxy (M 33), and it fits equally well or better than 
NFW for eleven galaxies (DDO 47, ESO 287-G13, IC 2574, NGC 3877, NGC 3917, NGC 3972, NGC 4085, NGC 7339, UGC 6399, UGC 6983, and UGC 11455).  In summary, the NFW's profile 
fits equally well or better for 16 galaxies (out of 30) and Burkert's profile achieves an equally well or a better fit for 26 galaxies (out of 30) than the standard Rindler model.  

\item The power-law generalized Rindler model ($n$-free) fits worst than Burkert's profile, but slightly better than NFW. This 
Rinder model fits equally well or better than both NFW and Burkert models for six 
galaxies (M 31, M 33, NGC 3953, NGC 4183, NGC 6503, and UGC 6917) and, in addition, it fits equally well or better than NFW for fourteen      
galaxies (DDO 47, ESO 287-G13, IC 2574, NGC 3877, NGC 3917,  NGC 3949, NGC 3972, NGC 4085, NGC 4217, NGC 7339, UGC 128, UGC 6399, 
UGC 6983, and UGC 11455). In summary, the NFW profile fits equally well or better for 14 galaxies (out of 30)  and 
Burkert's profile achieves an equally well or better fit for 25 galaxies (out of 30) than the generalized Rindler model.    The fact that Burkert's shallow profile fits better than the cuspy NFW 
profile to some of these galaxies has been reported in the literature \cite{Gentile:2004tb}, as well as further analysis on the NFW 
fits in Ref. \cite{Martins:2006wy}.  
\end{itemize}

\section{Conclusions} \label{grumiller:conclu} 
Using a collection of rotation curves of thirty galaxies, we have tested the standard ($n=1$) and power-law generalized ($n$-free) Grumiller's model 
of modified gravity at large distances.   The corresponding gravitational potential implies a new (Rindler) acceleration constant in nature that affects 
the rotation curve as $v_{T}^2 (r) = \gs v^2_{\star} + v^2_{G} + a |\vec{r}|^n $, where the last term would replace the contribution of the dark matter profile. 

The results of the fits are in Tables \ref{grumiller1} and  \ref{grumiller2}, and a comparison of the goodness-of-fit to NFW's and Burkert's profiles is presented in 
Table \ref{chi2_comp}.   Our results show that: i) the standard Rindler model ($n=1$) does not achieve good fits since only thirteen (of thirty) galaxies had $\chi^2_{\rm red} \le 1$, and    
these best-fitted galaxies also show a big spread in the acceleration parameter; 
ii)  the power-law generalized model ($n \neq 1$) does achieve an equally well (14/30) or a better fit (16/30) than the standard Rindler's model for all galaxies, but again only thirteen (of 30) galaxies had $\chi^2_{\rm red} \le 1$, and also these best-fitted galaxies show big spreads in the Rindler parameters; 
iii) the comparison of these modified gravity models with standard dark matter profiles yields that the standard Rindler model ($n=1$) fits worst Burkert's profiles, but it fits equally well as 
NFW. The generalized model achieves better fits than NFW's profile, but much poorer fits than Burkert's profile.  

The main problem, however, is that both Rindler parameters $(a,n)$ show at least one order of magnitude spread that cannot be 
explained by the corresponding uncertainties, not pointing to single universal values.  

In comparison with previous, similar studies \cite{Mastache:2012ep}, where seventeen THINGS galaxies were employed, the results 
here are less conclusive, since the spreads on the computed Rinder parameters are smaller. Nevertheless,  our present work points again to  
inconsistent standard and power-law generalized Rindler models.  

\section*{Acknowledgments}
We thank  P. Salucci for providing us with the rotation curves data used here.  We also thank Jorge Mastache for useful discussions.
JAGL acknowledges financial support from ININ.

\begin{table}[t]
\begin{center}
\begin{tabular}{  c | c   c  c  c  } \hline \hline
 & &   $\chi_{\rm red}^2$ & &  \\ \hline 
\bf Galaxy  & Burkert & NFW & ($n=1$) &  ($n$-free) \\ \hline 
 DDO 47	              &   $1.0$&$ 5.7 $& 4.9&1.2\\ \hline
ESO 116-G12	&  $0.9$&$ 2.6$&3.8&3.7\\ \hline
ESO 287-G13   &  $1.5$&$ 2.1$&1.8&1.6\\ \hline
IC 2574	         & $2.0$&$ 43.5$&38.9&2.4\\ \hline
M 31		         &  $1.9$&$ 1.7$&1.6&1.1\\ \hline
M 33		          &  $5.5	$&$4.1 $&4.3&3.3\\ \hline
NGC 55	         &    $	0.3$&$ 2.9$&3.7&3.5\\ \hline
NGC 300	         &     $0.4	$&$ 0.6$&0.7&0.7\\ \hline
NGC 1090         & $0.8	$&$ 1.8$&3.8&3.4\\ \hline
NGC 2403         &     $1.5	$&$1.1 $&4.0&2.3\\ \hline
NGC 3877         &   $0.3	$&$1.1 $&0.9&	0.9\\ \hline
NGC 3917          & $	0.7$&$ 3.9$&3.6&	3.4\\ \hline
NGC 3949          &$	0.2$&$ 0.6$&0.4&	0.4\\ \hline
NGC 3953	 &$	0.5$&$ 0.5$&0.4&	0.4\\ \hline
NGC 3972	  &$	0.1$&$ 0.6$&0.5&0.5\\ \hline
NGC 4085	  &   $	0.5$&$ 2.4$&2.0&2.0\\ \hline
NGC 4100	 &$0.6	$&$ 0.9$&1.0&1.0\\ \hline
NGC 4157          &$	0.5$&$ 0.6$&0.8&0.7\\ \hline
NGC 4183	 &$0.1	$&$ 0.1$&0.1&0.1\\ \hline
NGC 4217          &$0.3	$&$0.7 $&0.7&0.7\\ \hline
NGC 5585	 &$	0.4$&$ 4.9$&7.3&6.9\\ \hline
NGC 6503          &$2.1	$&$ 5.3$&6.8&1.4\\ \hline
NGC 7339          & $1.2$ & $1.5$&$1.4$ & $1.4$\\ \hline
UGC 128	           &$	0.0$&$ 0.2$&0.2&0.2\\ \hline
UGC 6399	  &$	0.0$&$ 0.4$&0.3&0.3\\ \hline
UGC 6917	  &$	0.4$&$ 0.3$&0.2&0.2\\ \hline
UGC 6983	 &$	0.6$&$0.7 $&0.7&0.7\\ \hline
UGC 8017	  &$	3.1$&$ 3.7$&6.3&$4.8$\\ \hline
UGC 10981	 &$	7.2$&$ 6.6$&16.0&10.4\\ \hline
UGC 11455    &$13.0 $&$ 19.3$&17.7&16.9\\ \hline \hline

\end{tabular} 
\end{center}
\caption{Summary of the $\chi_{\rm red}^2$ values for the different dark matter profiles and gravity models.} \label{chi2_comp} 
\end{table}


\end{document}